\pdfoutput=1
\documentclass[aps,prb,amsmath,amssymb]{revtex4-1}
\usepackage{graphicx}
\usepackage{dcolumn}
\usepackage{epstopdf}
\usepackage{bm}

\begin{document}

\title[A first-principles study of pyroelectricity in GaN and ZnO]{A first-principles study of pyroelectricity in GaN and ZnO}

\author{Jian Liu}
\email{Jian.Liu@stonybrook.edu}
\author{Maria V. Fern\'{a}ndez-Serra}
\author{Philip B. Allen}
\affiliation{Department of Physics and Astronomy, Stony Brook University, Stony Brook, NY 11794-3800, United States.}
\date{\today}

\begin{abstract}
First-principles calculations are made for the primary pyroelectric coefficients of wurtzite GaN and ZnO.  The pyroelectricity is attributed to the quasiharmonic thermal shifts of internal strains (internal displacements of cations and anions carrying their Born effective charges). The primary (zero-external-strain) pyroelectricity dominates at low temperatures, while the secondary pyroelectricity (the correction from external  thermal strains) becomes comparable with the primary pyroelectricity at high temperatures. Contributions from the acoustic and the optical phonon modes to the primary pyroelectric coefficient are only moderately well described by the corresponding Debye function and Einstein function respectively.

\begin{description}
\item[PACS numbers]
\end{description}
\end{abstract}

\pacs{Valid PACS appear here}

\maketitle

\section{Introduction}

Pyroelectricity $\vec{p}(T)$, defined as  temperature variation of the spontaneous polarization $\vec{P}_s$, is a fundamental and poorly understood property\cite{Resta,SBLang}. Among various applications, pyroelectric materials are widely used in thermal infrared (IR) detectors\cite{physicstoday} for their sensitivity over a wide range of temperatures. Among the non-ferroelectric pyroelectrics, wurtzite crystals exhibit spontaneous polarization and pyroelectricity comparable to ferroelectric pyroelectrics, and are candidates for high-temperature IR detection, because they do not have a Curie temperature at which the spontaneous polarization can be lost.\\


Crystals have specific free parameters that can vary without altering symmetry.  These are external strains and internal strains.  The external strains are components of the strain tensor $\epsilon_{\alpha\beta}$ that have full crystalline symmetry ($\Delta V/V$ if cubic, or $\Delta a/a$ and $\Delta c/c$ if hexagonal like wurtzite.)  The external strains will be denoted $\epsilon_i$.  The internal strains describe degrees of freedom of atoms in the unit cell.  An example is the $c$-axis cation-anion spacing denoted $uc$ in wurtzite, where $u$ is typically close to the ``ideal'' value 3/8 of perfect stacked tetrahedra, a value not required by symmetry.  The internal strains will be denoted $u_i$.  Wurtzite is the highest symmetry structure that can have spontaneous polarization, and has the minimal number of 2 external strains and 1 internal strain.  The polarization is strongly affected by the internal strain\cite{int_para} $u$, and the pyroelectricity is closely related to its temperature shift $du/dT$.\\

It is conventional to separate the total (at constant stress $\sigma$) pyroelectric coefficient $p_\sigma(T)$ 
into two parts\cite{Born,Szigeti}: the primary (at constant strain $\epsilon$) $p_\epsilon(T)$, and the secondary $p_2(T)$
\begin{equation}
p_{\sigma}(T)=\left(\frac{dP_s}{dT}\right)_{\sigma}=\left(\frac{\partial P_s}{\partial T}\right)_{\epsilon}+\sum\limits_i\left(\frac{\partial P_s}{\partial \epsilon_i}\right)_{T}\left(\frac{\partial \epsilon_i}{\partial T}\right)_{\sigma}=p_\epsilon(T)+p_2(T).
\end{equation}
Here we simplify the notation by assuming that polarization $\vec{P}=P\hat{z}$ occurs along a unique axis.  The label $z$ for this axis is dropped when unnecessary.
The primary part, $p_\epsilon(T)$, is the ``clamped-lattice'' pyroelectricity, where external strains are held fixed, but internal strains relax thermally.   The secondary part $p_2(T)$ accounts for  the changes that occur when external strains are allowed to develop.  \\

\section{Elementary Theory}


Harmonic vibrational normal modes are labeled by $(\vec{q}\lambda)$, wavevector and branch index.  The $\vec{q}=0$ optic modes of $A_1$ symmetry (invariant under all point-group operations), labeled $(\vec{0}j)$, are dynamic versions of the internal strains $u_j$.  Lattice anharmonicity allows the amplitudes $Q_{\vec{0}j}$ (see Appendix, Eq. (A3)) to develop static thermal internal strains $\langle Q_{\vec{0}j}(T)\rangle$.   This is one source of pyroelectric thermal shifts of $P_s$.  The other normal modes $\vec{q}\lambda$ have no allowed first-order static effect ($\langle Q_{\vec{q}\lambda}\rangle=0$), but their second-order static mean square amplitude $\langle Q_{\vec{q}\lambda}Q_{-\vec{q}\lambda}\rangle$ increases with $T$ in harmonic approximation.  These cause an additional electron-phonon source of thermal renormalization of $P_s$ even in the absence of internal and external strains. Both quasiharmonic internal strain and electron-phonon contributions to pyroelectricity are mentioned by Born \cite{Born} and Szigeti \cite{Szigeti}. After Szigeti's work, the electron-phonon part has been generally discounted as less important, and will be ignored in our work.
Then to first approximation, the temperature-dependent spontaneous polarization $P_s(T)$ varies linearly with internal strain.  For the primary term, this is
\begin{equation}
P_{s,\epsilon}(T)=P_{s,\epsilon}(0)+\sum\limits_{j}\frac{\partial P_{s,\epsilon}}{\partial Q_{\vec{0}j}} \langle Q_{\vec{0}j}\rangle.
\end{equation}
The sum goes over all the ``active'' $\vec{q}j=\vec{0}j$ phonons ($A_1$ modes).  In wurtzite, the  one relevant $A_1$ mode has opposite displacements $\vec{u}_{\kappa z}$ of anions and cations (labeled by $\kappa$) along the polar $c$ or $\hat{z}$ axis. 
The connection between static displacement $\langle\vec{u}_{\kappa z}\rangle$ of atom $\kappa$ in each cell, and normal mode amplitude $\langle Q_{\vec{0}j}\rangle$ is
\begin{equation}
\langle u_{\kappa z}\rangle=\sum_j \langle Q_{\vec{0}j}\rangle \epsilon_{\vec{0}j}(\kappa z)/\sqrt{M_\kappa},
\label{eq:}
\end{equation}
where $ \epsilon_{\vec{0}j}(\kappa\alpha)$ is the normalized $\alpha$-Cartesian component of the $(\vec{0}j)$ eigenvector of the usual (mass-weighted) harmonic dynamical matrix.
The dependence of $P_s$ on the internal displacement defines the ``Born effective charge'' $Z^\ast$, 
\begin{equation}
\frac{eZ_{\kappa}^{\alpha\beta}}{\Omega}=\frac{\partial P_{s,\epsilon}^{\alpha}}{\partial u_{\kappa\beta}},
\label{eq:}
\end{equation}
where $\Omega$ is the unit-cell volume.
 The magnitude of $Z^\ast$ governs the zone-center LO/TO splitting\cite{DFPT}.  The primary pyroelectric coefficient is given by
\begin{equation}
p_{\epsilon}^{\beta}(T) = \frac{e} {\Omega} \sum_{j,\kappa\alpha} Z_\kappa^{\beta\alpha} \frac{du_{\kappa\alpha}(\vec{0}j)} {dT}.
\label{eq:2terms}
\end{equation}
%
%
This ignores the electron-phonon term. In wurtzite, it simplifies to $p_{\epsilon}(T)=(2e/\Omega)Z^\ast d(uc)/dT$, where  $Z^\ast$ is the Born effective charge of the cation (the anion's is opposite by definition), and the factor of 2 recognizes the two molecules per unit cell.  Further details are given in the appendix.\\

\section{Notions and Evidence}

Born \cite{Born} and Szigeti \cite{Szigeti} present different-looking formulas of the temperature shift in Eq.(\ref{eq:2terms}). We find that they are equivalent.  An interesting experiment on wurtzite ZnO by Albertsson {\it et al.} \cite{ZnO-pc} 
measures the internal parameter shift $du/dT$ directly. They find that Eq.(\ref{eq:2terms}) matches $p_{\sigma}(T)$ provided $Z^\ast = 0.2$ is used. We believe that they have mis-defined $Z^\ast$ and that the correct definition makes the empirical $Z^\ast$ larger by 4, or $Z^\ast = 0.8$. Our results presented below are the first microscopic calculations of thermal shift of internal parameters.  Our results for $du/dT$ are smaller than the Albertsson experiment by $\approx 2$, and our computed $Z^\ast = 2.2$ is larger (agreeing with all modern calculations). We are not able to identify the source of the discrepancy, but our results also indicate that Eq.(\ref{eq:2terms}) is satisfactory.\\

Recent developments include the measurement of significant pyroelectricity of $c$-plane GaN at room temperature\cite{GaN-exp1,GaN-exp2}. Peng and Cohen\cite{LiNbO3} studied the origin of pyroelectricity in LiNbO$_3$ using molecular dynamics with a first-principles-based shell model potential. They found that the primary pyroelectric effect is the major part of the pyroelectricity, and comes from the anharmonic atomic displacement of participating ions carrying Born effective charges. This agrees with the estimate of Zook and Liu\cite{thinfilm} that the effects of clamping are negligible for the ferroelectric pyroelectrics. However they estimate a more significant secondary effect for the non-ferroelectric wurtzite pyroelectrics.  Spontaneous polarization at $T=0$ can now be predicted at the first-principles level\cite{Vanderbilt}.  However, predictions for pyroelectricity have not yet reached ``first-principles level''. Here we give a first-principles quasi-harmonic theory for pyroelectricity in wurtzite GaN and ZnO.\\

\section{Computational Method}
 
Following Szigeti \cite{Szigeti}, as derived in the appendix, the primary pyroelectric coefficient reads
%
\begin{equation}
p_{\epsilon}^{\beta}(T)=\sum\limits_{\kappa\alpha}\sum\limits_{\vec{0}j}\sum\limits_{\vec{q}\lambda}\frac{eZ^{\beta\alpha}_{\kappa}}{\Omega}\frac{2}{\hbar\omega_{\vec{0}j}}\sqrt{\frac{\hbar}{2M_{\kappa}\omega_{\vec{0}j}}}\epsilon_{\kappa\alpha}(\vec{0}j)V_3\left(\begin{array}{ccc}\vec{0}&\vec{q}&-\vec{q}\\j&\lambda&\lambda\end{array}\right)\frac{\partial (2n_{\vec{q}\lambda}+1)}{\partial T}.
\label{eq:panh}
\end{equation}
Here $\beta$ labels the direction of the spontaneous polarization, and $V_3$ is the third-order anharmonic coefficient for the active mode $\vec{0}j$ (see Appendix).  The sum on $\vec{q}\lambda$ runs over all phonon branches in the Brillouin zone. The Appendix shows that the anharmonic $V_3$ coefficient is related to ``internal'' Gr$\ddot{\rm u}$neisen parameters defined as $\gamma_{\vec{q}\lambda}(\vec{0}j)=-d\log\omega_{\vec{q}\lambda}/d\log Q_{\vec{0}j}$. These measure the shift of phonon frequency $\omega_{\vec{q}\lambda}$ per unit change in the amplitude $Q_{\vec{0}j}$ of the active modes.  They have been defined previously by Gibbons\cite{Gibbons}.  This part of the theory ignores the influence of external strains (the ``secondary'' effect), which will be added later using measured external strains $\epsilon_i(T)$ and computed piezoelectric coefficients \cite{LiNbO3,thinfilm}.  \\

In wurtzite structure, the active $A_1$ mode is split.  When $\vec{q}$ approaches 0 along the $c$ or $\hat{z}$ axis, it is a high frequency longitudinal branch denoted $A_1$(LO).  When $\vec{q}$ approaches 0 along lines in the $xy$ plane, it is an intermediate frequency transverse branch labeled $A_1$(TO). The difference, $\omega^2$($A_1$-TO)$-$$\omega^2$($A_1$-LO), comes from the long-range E-field of the LO polar vibration\cite{Gonze}. The frequency $\omega_{\vec{0}j}$ in the denominator of Eq.(\ref{eq:panh}) contains the Born-Oppenheimer restoring force restraining the thermal internal stress.  The rule is to use the TO frequency, which corresponds to a pyroelectric distortion in zero electric field.\\

Electronic structure calculations are performed using the {\sc Quantum ESPRESSO} package\cite{QE} within the local density
approximation (LDA)\cite{LDA}. We use norm-conserving Troullier-Martins pseudopotentials\cite{TM} in our calculations. The electronic wave-functions are expanded in a plane-wave basis with a kinetic energy cutoff of 180 Ry. Ga-3$d$ and Zn-3$d$ states are treated explicitly as valence states. We use a $6\times6\times4$ $k$-point mesh for Brillouin-zone sampling. Phonons are calculated using density-functional perturbation theory (DFPT)\cite{DFPT}. The third-order anharmonic coefficients $V_3(0j,\vec{q}\lambda,-\vec{q}\lambda')$ are computed on an $8\times8\times6$ $q$-point mesh through the finite difference of the dynamical matrix by displacing atoms along the displacement pattern $u_{\kappa\alpha}(\vec{0}j)$. The quasiharmonic internal shift $\langle Q_{\vec{0}j}\rangle$, derived from the ``internal'' Gr$\ddot{\rm u}$neisen parameter $\gamma_{\vec{q}\lambda}(\vec{0}j)$, involves only diagonal components ($\vec{q}\lambda=-\vec{q}\lambda'$). This is derived in the appendix, Eq. (A13).\\

\section{Results and Discussion}

\begin{figure}[htb!]
\centering
\includegraphics[scale=0.25]{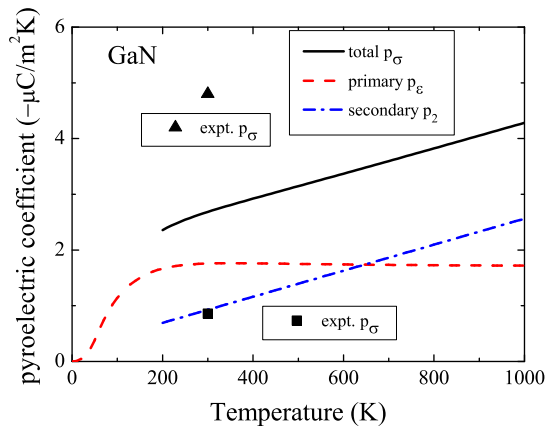}
\caption{The pyroelectric coefficient of GaN. Experimental values are from Ref. [\onlinecite{GaN-exp1}] (triangle) and Ref. [\onlinecite{GaN-exp2}] (square).}
\label{fig:1}
\end{figure}

\begin{figure}[htb!]
\centering
\includegraphics[scale=0.25]{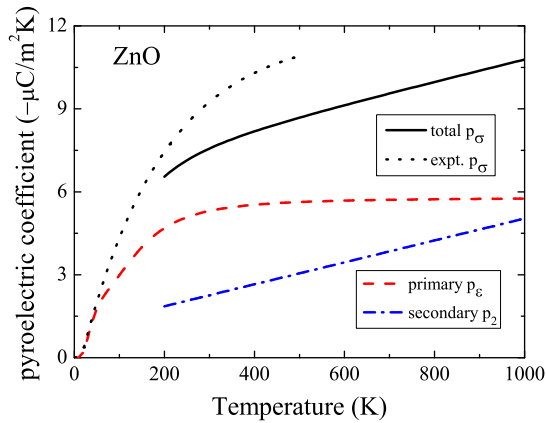}
\caption{The pyroelectric coefficient of ZnO. Experimental values are from Ref. [\onlinecite{Ibach}].}
\label{fig:2}
\end{figure}

Computed properties of GaN and ZnO are summarized in Table \uppercase\expandafter{\romannumeral1}.  In Figs. \ref{fig:1}-\ref{fig:2} we show the calculated primary pyroelectric coefficients $p_\epsilon(T)$ and the experimental total pyroelectric coefficient $p_\sigma(T)$ for GaN\cite{GaN-exp1,GaN-exp2} and ZnO\cite{Ibach} respectively. The secondary pyroelectric coefficients $p_2(T)$ are calculated from $2e_{31}\alpha_1+e_{33}\alpha_3$ using the measured linear thermal expansion coefficients $\alpha_1,\alpha_3$\cite{TEC} and the calculated piezoelectric stress constants $e_{31},e_{33}$\cite{Vanderbilt}. However, it is reported that for GaN and ZnO the computed piezoelectric constants are uncertain by as much as 30\%\cite{piezo1,piezo2}. Therefore the calculated $p_2(T)$ should be considered rough estimates. From Eq.(\ref{eq:panh}), it is clear that $p_\epsilon(T)$ follows the form of specfic heat. Therefore $p_\epsilon(T)$ vanishes as $T^3$ at low temperatures and saturates at high temperatures. Above room temperature, the secondary pyroelectric effect is comparable with the primary effect. This differs from ferroelectric pyroelectrics, where the primary pyroelectricity dominates\cite{LiNbO3}. For GaN, disagreement in the experimentally measured pyroelectric coefficients is reported\cite{GaN-exp1,GaN-exp2}, possibly due to the piezoelectric contribution from the strain introduced by the substrates. For ZnO, our calculated total pyroelectricity is lower than the experimental data, indicating the possible contribution from the electron-phonon effect, which is left out in our first-principles calculations.\\
\begin{table*}
\caption{\label{arttype} The calculated lattice constants, Born effective charge and long-wavelength $A_1$(TO) phonon frequency for GaN and ZnO. Experimental values are shown in parentheses except for Born effective charge where theoretical values are shown instead.}
\begin{ruledtabular}
\begin{tabular}{@{}ccccc}
&$a$(\AA)&$c$(\AA)&$Z^{33}$($e$)&$\omega_{TO}$(cm$^{-1}$)\\
\colrule
GaN&3.182 (3.187\footnotemark[1])&5.189 (5.186\footnotemark[1])&2.77 (2.72\footnotemark[2])&534 (533.8\footnotemark[3])\\
ZnO&3.219 (3.25\footnotemark[1])&5.195 (5.207\footnotemark[1])&2.28 (2.11\footnotemark[2])&390 (378\footnotemark[4])\\
\end{tabular}
\end{ruledtabular}
\footnotetext[1]{Ref. [\onlinecite{TEC}], X-ray powder diffractometry at 300K.}
\footnotetext[2]{Ref. [\onlinecite{Vanderbilt}], first-principles calculations in the local density approximation.}
\footnotetext[3]{Ref. [\onlinecite{GaN-phonon}], Raman spectra at 6K.}
\footnotetext[4]{Ref. [\onlinecite{ZnO-phonon}], inelastic neutron scattering spectra at 10K.}
\end{table*}

\begin{figure}[htb!]
\centering
\includegraphics[scale=0.25]{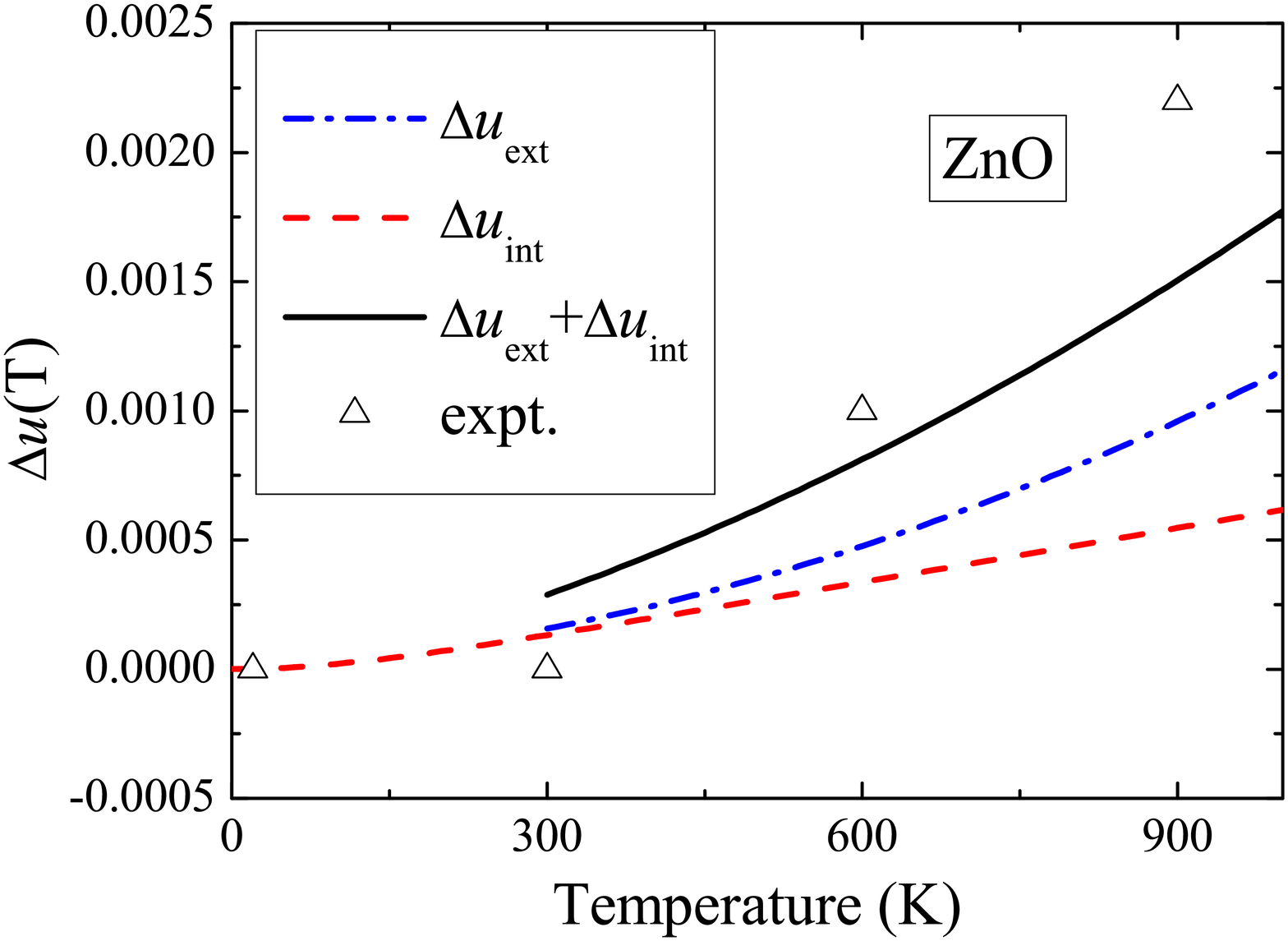}
\caption{Thermal shift of the ZnO internal parameter $u$.  The measured zero temperature value is 0.382 \cite{ZnO-u-1,ZnO-u-2}, close to the ``ideal'' value of 3/8. Experimental values are from Ref. [\onlinecite{ZnO-pc}].}
\label{fig:3}
\end{figure}

Figure \ref{fig:3} shows the predicted and the experimentally measured values of the internal parameter $u$ of ZnO.  The theory for pyroelectricity also generates a formula for the internal strain $u(T)$ which is closely parallel to the Gr\"uneisen quasiharmonic theory of volume expansion\cite{Gibbons},
\begin{equation}
\frac{\Delta u}{u}=\frac{1}{2M_{\rm red}\omega_0^2 c^2 u^2} \sum_{\vec{q}j} \left(n_{\vec{q}j}+\frac{1}{2}\right)\hbar\omega_{\vec{q}j}\gamma_{\vec{q}j}(0),
\label{eq:intexp}
\end{equation}
where the label 0 on $\omega_0$ and on the internal Gr\"uneisen parameter $\gamma_{\vec{q}j}(0)$ indicates the $A_1$(TO) mode.  This formula gives only the part of $u(T)$ that occurs when external strains are absent.  The full result is
\begin{equation}
u(T) = u(0) + [\Delta u(T) - \Delta u(0) ] +\left( \frac{\partial u}{\partial a}\right)_{\rm BO} [ a(T)-a(0)] + 
\left(\frac{\partial u}{\partial c}\right)_{\rm BO} [c(T)-c(0)].
\label{eq:uT}
\end{equation}
The value $u(0)$ from experiment contains all zero-point shifts.  The factor $[\Delta u(T) - \Delta u(0) ] $ comes from the theory of Eq.(\ref{eq:intexp}), and the factors $[ a(T)-a(0)] $ and $[ c(T)-c(0)] $ come from experiment \cite{TEC}.  For ZnO, the theoretical values of $\partial u/\partial a$ and $\partial u/\partial c$ are $0.083 \AA^{-1}$ and $-0.051 \AA^{-1}$ respectively, coming from our DFT Born-Oppenheimer calculations. In Fig. \ref{fig:3} we show for ZnO the thermal shift of the internal parameter $\Delta u(T)$. Our calculated thermal displacement increases monotonically with increasing temperature, while experimentally $u(T)$ remains unchanged between 20 and 300 K. Except for this discrepancy at low-$T$, the overall  agreement is satisfactory.\\

\begin{figure}[htb!]
\centering
\includegraphics[scale=0.25]{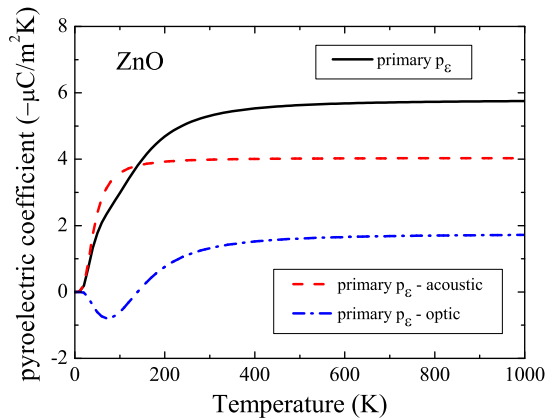}
\caption{The primary pyroelectric coefficient of ZnO: acoustic and optic branches.}
\label{fig:4}
\end{figure}

\begin{figure}[htb!]
\centering
\includegraphics[scale=0.25]{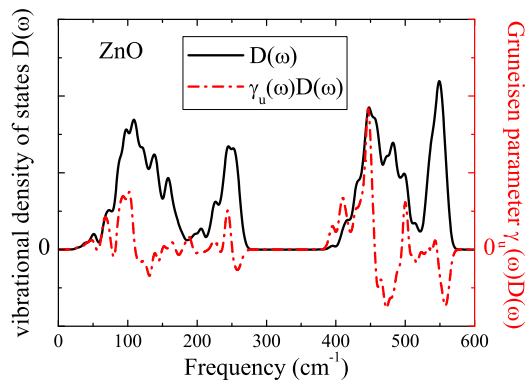}
\caption{Vibrational density of states $D(\omega)$ and internal gr$\ddot{\rm u}$neisen parameter $\gamma_u$($\omega$) of ZnO.}
\label{fig:5}
\end{figure}

Yan $et$ $al.$\cite{GaN-calc} conjecture a  temperature dependence of the primary pyroelectric coefficient $p_\epsilon(T)$ of GaN as a sum of Debye and Einstein functions. In Fig. \ref{fig:4} we show for ZnO our calculated contributions from acoustic and optic branches respectively. At low temperatures, only the acoustic phonon modes are sufficiently excited, while at high temperatures, contributions from the optic phonon modes become important. Our calculations indicate that for wurtzite ZnO, contributions from acoustic and optic branches are more complicated than Debye and Einstein functions, especially at low temperatures. In Fig. \ref{fig:5}, we show for ZnO the vibrational density of states $D(\omega)$, together with the frequency-distributed internal Gr$\ddot{\rm u}$neisen parameter $\gamma_u(\omega)$ defined as 
\begin{equation}
\gamma_u(\omega)D(\omega)=\sum\limits_{\vec{q}\lambda}\gamma_{\vec{q}\lambda}(\vec{0}j)\delta(\omega-\omega_{\vec{q}\lambda}).
\label{eq:}
\end{equation}
As an example of the use of this definition, the pyroelectric coefficient of wurtzite materials, Eq.(\ref{eq:panh}), is
\begin{equation}
p(T)=\frac{eZ^\ast}{2M_{\rm red}\omega_0^2 cu}\iint_0^\infty d\omega D(\omega) \gamma_u(\omega)C(\omega),
\label{eq:}
\end{equation}
where $C(\omega)$ is the harmonic specific heat of a mode of frequency $\omega$, $\hbar\omega (dn/dT)/\Omega$.
The total contribution to $p(T)$ is a complicated mix of contributions of both signs from acoustic and optic branches.\\


%

\section{Conclusions}
In summary, we have calculated the primary pyroelectric coefficients for wurtzite GaN and ZnO from first-principles. For wurtzite crystals the pyroelectricity was attributed to the anharmonic atomic displacements of the Born effective charges on the cations and anions. Good agreement was found between our first-principles calculations and the experimental data. We have shown that the primary pyroelectricity contributes the major part of the total pyroelectricity at low temperatures, while the secondary pyroelectricity becomes comparable with the primary pyroelectricity at high temperatures. The primary pyroelectric coefficient can be separated into contributions from acoustic and optic phonon modes, but these contributions can only moderately well described by Debye and Einstein functions respectively. The present study offers evidence that theory and computation can predict pyroelectricity with some reliability over a wide range of temperatures.

\appendix

\section{Primary pyroelectric effect: contribution from the anharmonic atomic displacement}
The temperature-dependent spontaneous polarization $P_s(T)$ can be expanded in terms of atomic displacement as
\begin{equation}
P_s(T)=P_s(0)+\sum\limits_{\vec{q}\lambda}\frac{\partial P_s}{\partial Q_{\vec{q}\lambda}}\langle Q_{\vec{q}\lambda}\rangle+\sum\limits_{\vec{q}\lambda\vec{q}'\lambda'}\frac{\partial^2 P_s}{\partial Q_{\vec{q}\lambda}\partial Q_{\vec{q}'\lambda'}}\langle Q_{\vec{q}\lambda}Q_{\vec{q}'\lambda'}\rangle,
\end{equation}
where $P_s(0)$ is the spontaneous polarization at $T=0K$. Under the rigid-ion approximation the second-order expansion term is neglected since the electron cloud follows the ion rigidly without deformation. The atomic displacement is written in terms of the phonon creation and annihilation operators,
\begin{equation}
u_{\kappa\alpha}^l=\sum\limits_{\vec{q}\lambda}\sqrt{\frac{\hbar}{2M_{\kappa}\omega_{\vec{q}\lambda}}}[\hat{a}_{\vec{q}\lambda}+\hat{a}^{+}_{-\vec{q}\lambda}]\epsilon_{\kappa\alpha}(\vec{q}\lambda)e^{i\vec{q}\cdot\vec{R}_l}.
\end{equation}
The connection between $u_{\kappa\alpha}^l$ and normal mode amplitude reads
\begin{equation}
u_{\kappa\alpha}^l=\sum\limits_{\vec{q}\lambda}\sqrt{\frac{1}{M_{\kappa}}}Q_{\vec{q}\lambda}\epsilon_{\vec{q}\lambda}(\kappa\alpha)e^{i\vec{q}\cdot l}.
\end{equation}
Only zone-center phonon terms are left after taking the thermodynamic average. The primary pyroelectric coefficient then reads
\begin{equation}
p_\beta(T)=\sum\limits_{j,\kappa\alpha}\frac{eZ^{\beta\alpha}_{\kappa}}{\Omega}\frac{\partial \langle{u_{\kappa\alpha}}(\vec{0}j)\rangle}{\partial T}.
\end{equation}
In order to evaluate the mean displacement, the potential energy is expanded in terms of atomic displacement to third order
\begin{equation}
V^{(3)}=\frac{1}{3!}\sum\limits_{\vec{q}\vec{q}'\vec{q}''jj'j''}V_3\left(\begin{array}{ccc}\vec{q}&\vec{q}'&\vec{q}''\\j&j'&j''\end{array}\right)(\hat{a}_{\vec{q}j}+\hat{a}^{+}_{-\vec{q}j})(\hat{a}_{\vec{q}'j'}+\hat{a}^{+}_{-\vec{q}'j'})(\hat{a}_{\vec{q}''j''}+\hat{a}^{+}_{-\vec{q}''j''}).
\end{equation}
Treating the cubic anharmonicity as a perturbation, the perturbed phonon wavefunction reads
\begin{equation}
\phi_{\tilde n}^{(1)}=\phi_{\tilde n}^{(0)}+\sum\limits_{\tilde n'}\frac{\langle \tilde n'^{(0)}|V^{(3)}|\tilde n^{(0)}\rangle}{E_{\tilde n}^{(0)}-E_{\tilde n'}^{(0)}}\phi_{\tilde n'}^{(0)}.
\end{equation}
The atomic displacement is then
\begin{equation}
\langle{Q_{\vec{0}j}}\rangle=\frac{2\sum\limits_{\tilde n}e^{-\beta(n+\frac{1}{2})\hbar\omega}\sum\limits_{\tilde n'}\frac{\langle \tilde n^{(0)}|Q_{\vec{0}j}|\tilde n'^{(0)}\rangle\langle \tilde n'^{(0)}|V^{(3)}|\tilde n^{(0)}\rangle}{E_{\tilde n}^{(0)}-E_{\tilde n'}^{(0)}}}{\sum\limits_{\tilde n}e^{-\beta(n+\frac{1}{2})\hbar\omega}}.
\end{equation}
The first Dirac bracket is non-zero only for $\tilde n'=\tilde n\pm1$. Therefore the second Dirac bracket reduces to terms containing $\hat{a}_{\vec{0}j}\hat{a}_{\vec{q}\lambda}\hat{a}^{+}_{-\vec{q}\lambda}$, $\hat{a}_{\vec{0}j}\hat{a}^{+}_{\vec{q}\lambda}\hat{a}_{-\vec{q}\lambda}$, $\hat{a}^{+}_{\vec{0}j}\hat{a}_{\vec{q}\lambda}\hat{a}^{+}_{-\vec{q}\lambda}$ and $\hat{a}^{+}_{\vec{0}j}\hat{a}^{+}_{\vec{q}\lambda}\hat{a}_{-\vec{q}\lambda}$.
\begin{equation}
\langle{Q_{\vec{0}j}}\rangle=-\sum\limits_{\vec{q}\lambda}\frac{2n_{\vec{q}\lambda}+1}{\hbar\omega_{\vec{0}j}}Q_{\vec{0}j}V_3\left(\begin{array}{ccc}\vec{0}&\vec{q}&-\vec{q}\\j&\lambda&\lambda\end{array}\right)
\end{equation}
More specifically,
\begin{equation}
\langle{u_{\kappa\alpha}(\vec{0}j)}\rangle=-\sum\limits_{\vec{q}\lambda}\frac{2n_{\vec{q}\lambda}+1}{\hbar\omega_{\vec{0}j}}\sqrt{\frac{\hbar}{2M_{\kappa}\omega_{\vec{0}j}}}\epsilon_{\kappa\alpha}(\vec{0}j)V_3\left(\begin{array}{ccc}\vec{0}&\vec{q}&-\vec{q}\\j&\lambda&\lambda\end{array}\right),
\end{equation}
where the anharmonic coefficient $V$ is given by the third derivative of the total energy with respect to the atomic displacement as
\begin{equation}
\begin{split}
V_3\left(\begin{array}{ccc}\vec{0}&\vec{q}&-\vec{q}\\j&\lambda&\lambda'\end{array}\right)=&\sum\limits_{\kappa_0\kappa_1\kappa_2,\alpha_0\alpha_1\alpha_2}\sqrt{\frac{\hbar^3}{8M_{\kappa_0}M_{\kappa_1}M_{\kappa_2}\omega_{\vec{0}j}\omega_{\vec{q}\lambda}\omega_{-\vec{q}\lambda'}}}\epsilon_{\kappa_0\alpha_0}(\vec{0}j)\epsilon_{\kappa_1\alpha_1}(\vec{q}\lambda)\epsilon_{\kappa_2\alpha_2}(-\vec{q}\lambda')\\&\times\left(\sum\limits_{l_1l_2}\frac{\partial^3 E}{\partial u_{\kappa_0\alpha_0}^{l_0}\partial u_{\kappa_1\alpha_1}^{l_1}\partial u_{\kappa_2\alpha_2}^{l_2}}e^{i\vec{q}\cdot\left(\tau_{l_1}-\tau_{l_2}\right)}\right).
\end{split}
\end{equation}
$V_3(\vec{0}j,\vec{q}\lambda,-\vec{q}\lambda')$ can also be obtained from the derivative of the dynamical matrix $D_{\alpha_1\alpha_2}(\kappa_1\kappa_2,\vec{q})$ with respect to the displacement pattern $Q_{\vec{0}j}$ as
\begin{equation}
\begin{split}
V_3\left(\begin{array}{ccc}\vec{0}&\vec{q}&-\vec{q}\\j&\lambda&\lambda'\end{array}\right)=&\sqrt{\frac{\hbar^3}{8\omega_{\vec{0}j}\omega_{\vec{q}\lambda}\omega_{-\vec{q}\lambda'}}}\sum\limits_{\kappa_1\kappa_2,\alpha_1\alpha_2}\epsilon_{\kappa_1\alpha_1}(\vec{q}\lambda)\epsilon_{\kappa_2\alpha_2}(-\vec{q}\lambda')\left(\frac{\partial}{\partial Q_{\vec{0}j}}\sum\limits_{l_1l_2}\frac{1}{\sqrt{M_{\kappa_1}M_{\kappa_2}}}\frac{\partial^2 E}{\partial u_{\kappa_1\alpha_1}^{l_1}\partial u_{\kappa_2\alpha_2}^{l_2}}e^{i\vec{q}\cdot\left(\tau_{l_1}-\tau_{l_2}\right)}\right).
\end{split}
\end{equation}
Through the diagonalization of the dynamical matrix $D_{\alpha_1\alpha_2}(\kappa_1\kappa_2,\vec{q})$ we have
\begin{equation}
\sum\limits_{\kappa_1\kappa_2,\alpha_1\alpha_2}\epsilon_{\kappa_1\alpha_1}(\vec{q}\lambda)\left(\sum\limits_{l_1l_2}\frac{1}{\sqrt{M_{\kappa_1}M_{\kappa_2}}}\frac{\partial^2 E}{\partial u_{\kappa_1\alpha_1}^{l_1}\partial u_{\kappa_2\alpha_2}^{l_2}}e^{i\vec{q}\cdot\left(\tau_{l_1}-\tau_{l_2}\right)}\right)\epsilon_{\kappa_2\alpha_2}(-\vec{q}\lambda')=\omega_{\vec{q}\lambda}^2\delta_{\lambda\lambda'}.
\end{equation}
The relation between $V_3(0j,\vec{q}\lambda,-\vec{q}\lambda)$ and the ``internal'' Gr$\rm{\ddot u}$neisen parameter reads 
\begin{equation}
\begin{split}
V_3\left(\begin{array}{ccc}\vec{0}&\vec{q}&-\vec{q}\\j&\lambda&\lambda\end{array}\right)=&\sqrt{\frac{\hbar^3}{8\omega_{\vec{0}j}\omega_{\vec{q}\lambda}\omega_{-\vec{q}\lambda}}}\frac{\partial \omega_{\vec{q}\lambda}^2}{\partial Q_{\vec{0}j}}=-\left(\frac{\hbar}{2\omega_{\vec{0}j}}\right)^{1/2}\frac{\hbar\omega_{\vec{q}\lambda}}{2Q_{\vec{0}j}}\gamma_{\vec{q}\lambda}(\vec{0}j),
\end{split}
\end{equation}
where the internal Gr\"uneisen parameter is defined as
\begin{equation}
\gamma_{\vec{q}\lambda}(\vec{0}j)=-\frac{Q_{\vec{0}j}}{\omega_{\vec{q}\lambda}}\frac{\partial \omega_{\vec{q}\lambda}}{\partial Q_{\vec{0}j}}.
\end{equation}
Combining (A8), (A13) and (A14), the temperature-dependent atomic displacement reduces to
\begin{equation}
\langle{Q_{\vec{0}j}}\rangle=-\sum\limits_{\vec{q}\lambda}\frac{\hbar}{2}\frac{2n_{\vec{q}\lambda}+1}{\omega_{\vec{0}j}^2}\frac{\partial \omega_{\vec{q}\lambda}}{\partial Q_{\vec{0}j}}.
\end{equation}
Here is an alternative derivation of Eq. (A15). Under the ``clamped-lattice'' condition, the Born-Oppenheimer potential energy is harmonic with respect to $u_{\kappa\alpha}(\vec{0}j)$:
\begin{equation}
U_{BO}=U_0+\frac{1}{2}\omega_{\vec{0}j}^2Q_{\vec{0}j}^2,
\end{equation}
where $Q_{\vec{0}j}$ is the normal coordinate $\sqrt{\sum\limits_{\kappa\alpha}M_{\kappa}u_{\kappa\alpha}^2(\vec{0}j)}$, and $u_{\kappa\alpha}$ is the atomic displacement of $\kappa$th atom in $\alpha$-direction. The Helmholtz free energy reads
\begin{equation}
F=U_{BO}+k_BT\sum\limits_{\vec{q}\lambda}\ln\left(2\sinh\frac{\hbar\omega_{\vec{q}\lambda}}{2k_BT}\right).
\end{equation}
The temperature-dependent atomic displacement $\langle u_{\kappa\alpha}(\vec{0}j)\rangle$ minimizes the Helmholtz free energy $F$. We then have 
\begin{equation}
\omega_{\vec{0}j}^2\langle Q_{\vec{0}j}\rangle=-\sum\limits_{\vec{q}\lambda}\frac{\hbar}{2}\left(2n_{\vec{q}\lambda}+1\right)\frac{\partial \omega_{\vec{q}\lambda}}{\partial Q_{\vec{0}j}}.
\end{equation}

\begin{acknowledgments}
We thank the Brookhaven National Laboratory Center for Functional Nanomaterials (CFN) under project 33862 for time on their computer cluster. This research also used computational resources at the Stony Brook University Institute for Advanced Computational Science (IACS). Work at Stony Brook was supported by US DOE Grant No. DE-FG02-08ER46550 (PBA) and DE-FG02-09ER16052 (MFS). Jian Liu is also sponsored by the China Scholarship Council (CSC).
\end{acknowledgments}


\nocite{*}

\bibliography{pyro}

\end{document}